\documentclass[10pt,twocolumn,letterpaper]{article}

\usepackage{times}
\usepackage{epsfig}
\usepackage{graphicx}
\usepackage{subfigure}
\usepackage{amsmath}
\usepackage{amssymb}
\usepackage{indentfirst}
\usepackage{booktabs}
 \usepackage[pagebackref=true,breaklinks=true,letterpaper=true,colorlinks,bookmarks=false]{hyperref}

\begin{document}

%%%%%%%%% TITLE
\title{Compressed Video Action Recognition with Refined Motion Vector}

\author{
\normalsize{Hanyuan~Cao*$^{1}$ \quad Shining~Yu*$^{2}$ \quad Jiashi~Feng$^{1}$ } \\
\small{$^{1}$ National Univeristy of Singapore \quad $^{2}$ Nanyang Technological University, Singapore} \\
\small{caohaoyuan@u.nus.edu \quad yush0012@e.ntu.edu.sg \quad elefjia@nus.edu.sg}\\
}

\maketitle
%\thispagestyle{empty}

%%%%%%%%% ABSTRACT
\begin{abstract} 
\let\thefootnote\relax\footnotetext{* indicates equal contribution.}

Although CNN has reached satisfactory performance in image-related tasks, using CNN to process videos is much more challenging due to the enormous size of raw video streams. In this work, we propose to use motion vectors and residuals from modern video compression techniques to effectively learn representation of the raw frames and greatly remove the temporal redundancy, giving faster video processing model. Compressed Video Action Recognition(CoViAR) has explored to directly use compressed video to train the deep neural network, where the motion vectors was utilized to present temporal information. However, motion vector is designed for minimizing video size where precious motion information are not obligatory. Compared with optical flow, motion vectors contains noisy and unreliable motion information. Inspired by the mechanism of video compression codecs, we propose an approach to refine the motion vectors where unreliable movement will be removed while temporal information is largely reserved. We prove that replacing the original motion vector with refined one and using the same network as CoViAR has achieved state-of-art performance on the UCF-101 and HMDB-51 with negligible efficiency degrades comparing with original CoViAR.

\end{abstract}

%%%%%%%%% BODY TEXT
\section{Introduction}  
Although CNN has obtained very high accuracies in image-related tasks such as image classifications and segmentations \cite{He2016image,Wang2016image,Hinton2012image}, its performance on video-related tasks, however, is far from satisfying. This is because CNN itself lacks an effective representation of the temporal information in videos. Some attempts to solve this problem is to integrate CNN together with RNN like LSTM\cite{Sriva2015LSTM,Xingjian2015LSTM,Yao2017lstm} to capture temporal information, or to build 3D CNNs\cite{Du20173DCNN,Ali20163DCNN,Joao20173DCNN,Tran20153DCNN,Varol20173DCNN} to include the temporal channel. However, these temporal representations are usually outperformed by two-stream methods which use a traditional algorithm - optical flow\cite{LK1981OF,Horn1981OF,Barron1994OF}, indicating that optical flow might be more effective in modelling motions. But an important drawback of optical flow is its high time complexity. As a result, there have been many works trying to train specific CNNs to approximate the computation of optical flows\cite{ESOF1,ESOF2,ESOF3}. However, these methods generally use very deep CNNs, which itself is still slow for non-GPU machines. On the other hand, another effective way of modelling temporal information without using optical flow is proposed recently\cite{CoViAR}, by directly working on compressed video domain and making use of information like motion vectors that are already stored in compressed videos. The comparable performances of optical-flows and compressed video action recognition indicate that there might be some intrinsic connections between the two techniques. But current optical flow algorithms all work on uncompressed RGB frames, and hence through this paper we develop an optical-flow-like feature that works directly on compressed videos, by approximating optical flows with motion vectors.

\section{Related Works}
\subsection{Action Recognition}
Before the thrive of CNNs, to handle the task of action recognition, hand-crafted features are extracted frist. Some famous features are Histogram of Oriented Gradients (HOG)\cite{HOG}, Histogram of Optical Flow (HOF)\cite{HOF} and improved Dense Trajectory (iDT)\cite{iDT}. Then these features are encoded into sparse or dense feature vectors and classified by a classifier.

With the development of CNN, the task of action recognition has been improved in the past few years. But this improvement is limited in that the CNN itself does not include methods to represent temporal information. To solve this problem, some attempts have been made to sample few frames in a video and merge classification results using pooling\cite{Simon201twostream}, or to combine CNN together with RNNs\cite{Yao2017lstm,Sriva2015LSTM,Xingjian2015LSTM}, but these attempts do not see significant improvements.

With years of research, current action recognition methods fall into two main categories: 3D CNNs and two-stream methods. Both of the methods could use temporal features extracted by optical flows as a stage of their pipelines to improve performances.
\begin{itemize}
\item 3D CNN\cite{Du20173DCNN,Ali20163DCNN,Joao20173DCNN,Tran20153DCNN,Varol20173DCNN}: use a 3D kernel for CNN so that it can perform temporal convolution to extract temporal info. The method could work alone, but when working together with optical flow features it see significant improvements on accuracy.
\item Two-Stream network\cite{Simon201twostream,Feich2017twostream}: The architecture consists of two CNN streams whose classification scores are fused at the end. One stream is called spatial stream, which is a CNN that accepts RGB inputs. The other stream is called temporal stream, which is a CNN that accepts optical flows as inputs. There many modifications of two-stream network, for example temporal segment network\cite{TSN}.
\end{itemize}

\begin{figure*}
\centering
\includegraphics[height=6cm, width=16cm]{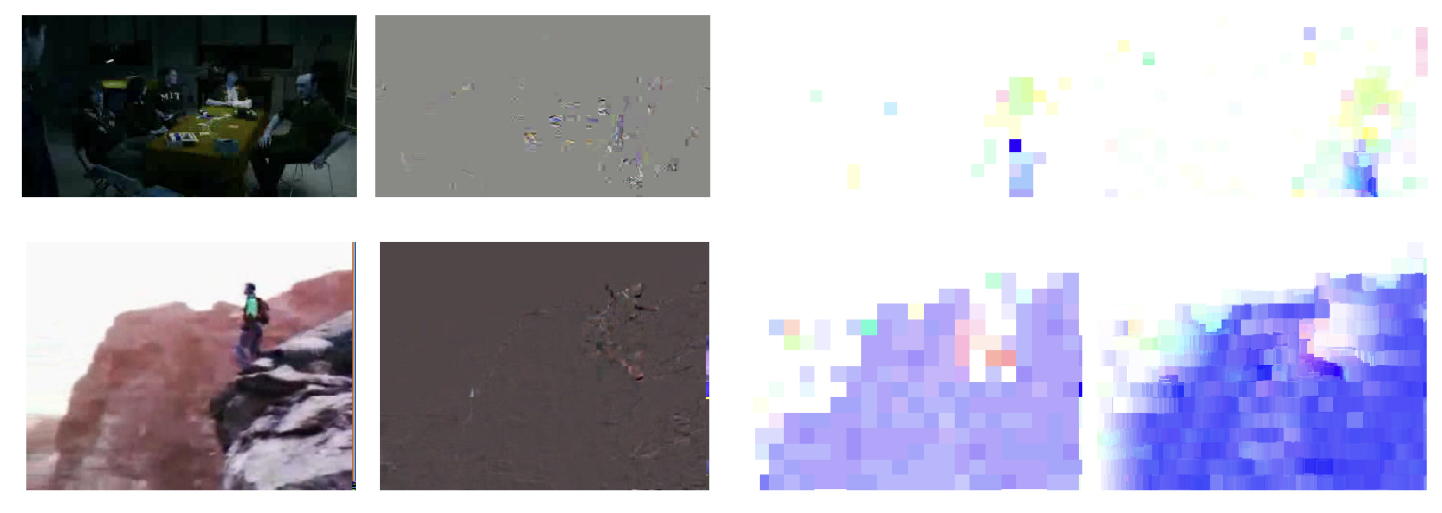}
\caption{Information Stored In Compressed Video \& Effect of accumulated motion vectors. Left to Right: Raw I-frame;Raw Residual;Raw Motion Vector; Accumulated Motion Vector. As shown in the figure 1, by up-sampling and accumulating motion vectors, smoother vector fields are obtained. Videos from Hmdb51 dataset.}
\label{fig:final_res}
\vspace{-.8em}
\end{figure*}

\subsection{Video Compression and Compressed Video Action Recognition}
Modern video compression techniques like MPEG-4\cite{MPEG1991}, H.264 minimize temporal redundancies between frames. Information are generally stored into three types of frames: I-frame, P-frame, and B-frame. I-frame is similar to an RGB image and is coded using only information from current frame. Between I-frames there are several P-frames. For each P-frame, only it’s change to the previous frame is stored. The stored information contains the ‘motion vector’ –  how blocks of pixels change with regarding to previous frame, and ‘residual’ – information needed to recover details not captured by motion vectors.%%% figure 1

Let $\Delta^{(t)}$ be the residuals and $\tau^{(t)}$ be the motion vectors at time \textit{t}, then P-frames can be recursively reconstructed using

\begin{equation}
I(i,t) = I(i-\Delta^{(t)}, t-1) + \tau^{(t)}
\end{equation}
where $I(i,t)$ denotes the pixel intensity at time $t$ for pixel $i$.

 Besides, B-frame is the type of frame with highest compressed rate. It is a bidirectional frame that can be referenced by both frames before and after it. But since action recognition pipeline is a casual system and should not reference future frame, B-frame is treated just as P-frame in previous work of compressed video action recognition\cite{CoViAR} (We refer to as Coviar in the following). 
For MPEG-4 format whose group of pictures is of size 12, if we treat B-frame as P-frame, there are on average 11 P-frame for each I-frame.

The work Coviar takes leverage of the information already stored in the compressed domain. For example, the motion vectors can be treated as pre-calculated optical flows. The work builds three CNNs flowing temporal segments models\cite{TSN}, one for I-frame, one for motion vectors stored in P-frame, and one for residuals stored in P-frame. The three networks are trained separately, and the prediction scores are fused at the end. To be specific, the CNNs are Resnet-152\cite{resnet} for I-frame, and Resnet-18\cite{resnet} for motion vectors and residuals to save computation effot. 

Another interpretation of the three subnets in Covair is that it also leverages from the two-stream architecture, with the I-frame subnet performing the role of RGB stream and mv subnet performing similarly to optical flow stream by capturing the temporal information in video stream. 

\subsection{Optical Flows and Approximate Optical Flows with Motion Vectors}

Optical Flow assumes that for a pixel at location $(x, y)$ at a frame at time $t$, its intensity should be the same as its intensity in the next frame at its new position $(x + \Delta x, y + \Delta y)$ at time $t + \Delta t$:
\begin{equation}
I(x,y,t) = I(x + \Delta x, y + \Delta y, t + \Delta t)
\end{equation}
The assumption is called the brightness constancy. By taking the Taylor series approximation of the right hand-side and divide the equation by $\Delta t$, we have 
\begin{equation}
\frac{\partial I}{\partial x} u_x + \frac{\partial I}{\partial y} u_y + \frac{\partial I}{\partial t} = 0
\end{equation}
where $u_x$ and $u_y$ denote the velocity of optical flow along $x$ and $y$ direction. There are lots of methods to determine the solutions $u_x$ and $u_y$ of the constraint\cite{LK1981OF,Barron1994OF}. However, these methods require lots of computational powers and space to save the results. And among these methods the most commonly used are the LK optical flow and dense optical flow, with good trade-off between computation costs and accuracy. There are also extensions to this classic algorithms to include confidence measure for the optical flows, such as Simoncelli's algorithm which use gradients to measure confidences for LK optical flow. 

Today optical flow still shows supreme performance in the task of action recognition. It could work together with RGB network to form the state-of-art two-stream architectures. And many researches today focus on approximating optical flow’s computation with CNNs. 

There have been attempts to approximate optical flows with motion information stored in compressed domain long before the invention of deep learning. In work Approximating Optical Flow Within the MPEG-2 Compressed Domain\cite{refineMV}, the authors proposed a confidence measure for motion vectors based on the inspiration of confidence measure for optical flows. For Simoncelli's algorithm, confidence of optical flows are measured by the first eigenvalue $\lambda_1$ of matrix $\overline{M}$: 

\begin{equation} 
\overline{M} =\sum_{i} M
\end{equation}
\begin{equation} M =
\begin{bmatrix}
    f_{x}^{2} & f_{x}f_{y} \\
    f_{x}f_{y} & f_{y}^{2}
\end{bmatrix}
\end{equation}
where $f_x = \frac{\partial I}{\partial x}$ and $f_y = \frac{\partial I}{\partial y}$. The obtained eigenvalue $\lambda_1$ is larger at edge areas than texture areas. Similarly,  the authors also used edge strengths (AC[1] and AC[8] of DCT Coefficients) as a confidence measure. The reason why strong edges signal confident motions could be explained by aperture problem, which says motions are hard to detect when there is no edge for reference.

Since the edge strength is only defined for I-frame, the author recursively compute the confidence maps for P-frames by using the confidence of the I-frame macroblock that each motion vector points from. The obtained confidence map is used to threshold low confidence motions that are likely to be noise. A smooth vector map that looks very similar to LK optical flows are achieved.  

Other researchers have tried to use this refined motion vectors to perform the task action recognition, with non-deep learning method. And it does see improvements in accuracy compared to working directly on raw motion vectors\cite{refineMV2}. 

\begin{figure*}
\centering
\includegraphics[height=7.5cm, width=14.5cm]{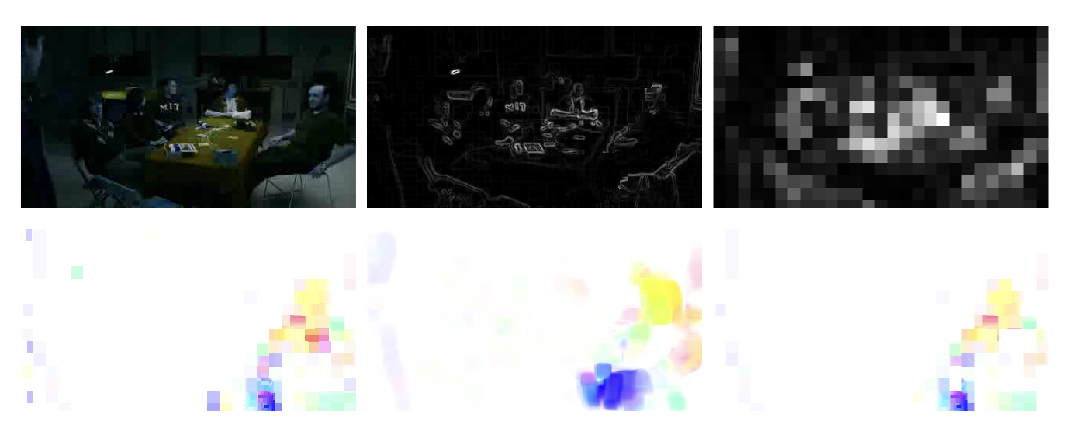}
\includegraphics[height=7.5cm, width=14.5cm]{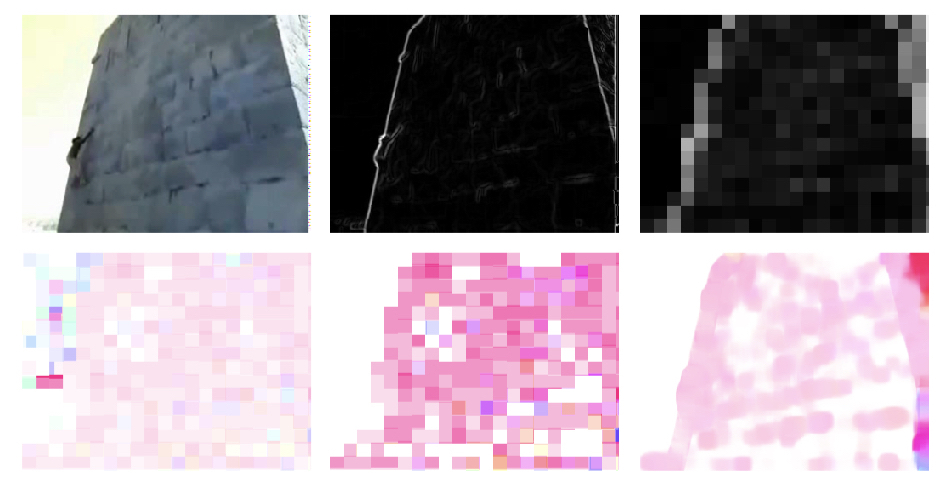}
\caption{Result of Our Proposed Method. Top Row Left to Right: Raw I-frame; Raw Motion Vector; Dense Optical Flow; Bottom Row Left to Right: Computed Pixel-Level Confidence Map; Computed Block-Level Confidence Map; Result of Our Proposed Method. Our result successfully removes the noises in raw motion vectors and obtained field are more similar to optical flow results.Videos from Hmdb51 dataset.}
\label{fig:final_res}
\vspace{-.8em}
\end{figure*}
\section{Compressed Video Action Recognition with Refined Motion Vectors}  %%%%%%%%%%%%%%%%%%%%% Chapter 3 %%%%%%%%%%%%%%%%

Rather than feeding the raw motion vectors into the Residual network, in this paper, we propose a process of refining motion vectors to approximate optical flows, and use the refined motion vectors to perform action recognition. The whole refinement process is as below:
\begin{itemize}
\item	For every P-frame selected for TSN (temporal segment network), compute edge strengths for its previous I-frame using Scharr filters and use it as the confidence map for the I-frame. 
\item	Extract the raw motion vectors from the compressed video domain. Up-sample motion vector fields to the same size of the I-frame, with a block of 16 * 16.  
\item	Recursively accumulate the motion vector of each pixel until the desired P-frame, the result constitutes a field indicating how to referencing each pixel in the P-frame to the previous I-frame. 
\item   Every pixel in P-frame is assigned a confidence value same as the referenced I-frame pixel's confidence value. The result constitutes a so called 'pixel-level' confidence map. 
\item	For the obtained ‘pixel-level’ confidence map, run a 3 * 3 median filter and compute average confidence for each 16 * 16 block, assigning the averaged confidence to the whole block to achieve ‘block-level’ confidence maps.
\item	Threshold the ‘block-level’ confidence maps with a fixed threshold of 0.0075 (for normalized confidence between [0,1]). Motion vectors of pixels with confidence below the threshold is set to zero. Others are reserved. The refined motion vectors are then passed to a Resnet18 classifier.
\end{itemize}
Details and reasons for this processes are explained below.

\textbf{Edge strength as motion confidence.} Following the idea of strong edges signals strong motion confidence. We utilize the well-established Sobel filters for computing the edge strength. However, since the edge strength could only be obtained for I-frames, we recursively trace the P-frame pixels to the I-frame pixels they point from to approximate the edge strength/motion confidence of the P-frame. 

\textbf{Up-sampling the motion vector fields.} Instead of using the original motion vector fields which is 16 * 16 smaller than I-frame, we up-sampled them back to the same size of I-frame. With this modification, as we recursively accumulate confidence back through the GOP, different ‘pixels’ in the same block in the up-scaled motion vector could point to different pixels in the original I-frame (otherwise there is a one-to-one relation between macroblocks in I-frame and motion vector fields), generating a smoother field which shows clearer motion pattern as shown in Figure 1. %%%%%%%%% figure1 

\textbf{Error propagation versus motion information sufficiency.} One key potential problem of the recursive computation of motion vector is the error propagation, i.e. if there is an error in an early frame of a GOP, the error will propagate and influence all frames in the later part and the motion vector will be inaccurate. To addressed issue, we attempted to directly use the I-frame's confidence map as all other P-frames' confidence map in that GOP. However, this attempt generate worse result than original Coviar, indicating that the removal of error propagation could not compensate the loss of motion information. 

\textbf{‘Pixel-level’ confidence map versus ‘Block-level’ confidence map.}
Originally, after obtaining the pixel-level confidence map for the motion vector field, the 16 * 16 average pooling is not performed after the 3 * 3 median filter. Though the result was better than Coviar' mv subnet, it does not outperform original Coviar when it works together with I-frame subnet. This result can be explained by the fact that resnet18 is too shallow to capture the pixel-level motion information. The typical two stream methods requires resnet152 to train the optical flow\cite{Feich2017twostream}. Besides, considerable useful motion vectors are aborted in this process, because the reliability of a block's motion vectors depends on whether their is a strong edge in that block. If there is, the whole block's motion vectors are reliable, vice versa. Therefore, the average pooling is introduced: Pixels whose edge strength is low but are in the same block with high-edge-strength pixels are still considered to be confident and are reserved, this is consistent with the motion vector's matching mechanism. By utilizing the block-level confidence map, reliable movement are reserved to the utmost extent while noises are effectively removed. %%%%% Figure 2

\textbf{Choosing the optimal threshold} We tested various adaptive threshold before the fixed threshold is eventually chosen. For example, we tried the strategy that only keep the motions with top 80\% confidence among all motions in a field. We validated a range of adaptive threshold on hmdb51 dataset, from 30\% to 90\%, and the optimal threshold was at 80\%, where both mv alone and mv + I-frame improved compared with original Coviar. After experimenting with different adaptive functions, the fixed value 0.8\% for normalized confidence map is determined to be the optimal. This means if the block's confidence value is lower than 0.8\% of the highest confidence value, all pixels' motion in this block will be set to zero.

\section{Experiments}
\subsection{Evaluation Dataset}
Two datasets are used for evaluation of our proposed method: Hmdb51\cite{HMDB51} and UCF101\cite{UCF101} action recognition datasets. They are the most commonly used metrics for action recognition performance measurement. Both datasets consists of three splits. For Hmdb51 each split consists around 3500 training videos and 1500 testing videos (in total 6766 clips), with 51 different action categories. Performances on Hmdb51 are generally worse than UCF101 dataset, because there are fewer number of training clips and video clips often contain changing of camera views or irrelevant actions.For UCF101 dataset, each split consists of around 7500 training videos and 3500 testing clips (in total 13320 clips), with 101 different categories. 

\subsection{Tranining Specification}
We trained and tested the model on an Ubuntu 16.04 machine, with Intel 8700K CPU, NVIDIA 12GB GTX 1080ti GPU and 16GB RAM. CUDA version is 8.0, with cudnn 6.0, and our codes are implemented using torch 0.4.1. The training process takes around 6 hours on a Hmdb51 split for the mv subnet, while test takes few minutes. The number of workers used is only one.
For efficiency consideration, refinements of motion vector discussed in Section 3 are implemented with Cython.

For the training process, we used Adam optimizer, with epsilon of 0.001 and no weight decay. For the Hmdb51 dataset, the starting learning rate is 0.05 and training takes 360 epochs in total. The learning rate decays to 10\% at epoch 120, epoch 200, epoch 280. For the UCF101 dataset, the starting learning rate is 0.01 and training takes  510 epochs in total. The learning rate decays 0.1 at epoch 120, epoch 200, epoch 280. The learning rate decays to 10\% at epoch 150, epoch 270, epoch 390.

\subsection{Accuracy}
\label{sec:4.3}
The performance comparison of the proposed method and original Coviar's mv models are shown in Table~\ref{tab:tab1}. Our proposed refined mv generally slightly improves the accuracy.
 \begin{table}[h]
\begin{center}
 \caption{Accuracy Comparison MV Alone}
 \label{tab:tab1}
 \begin{tabular}{c c c c c} 
 \hline
& Split1 & Split2 & Split3 & Average \\ 
 \hline
 \textbf{Hmdb51} & & & & \\ 
 Coviar mv & 40.78 & \textbf{40.33} & 41.18 & 40.76\\ 
 
 Refined mv& \textbf{41.63} & 39.02 & \textbf{42.81} & \textbf{41.15}\\
 
  \textbf{UCF101} &&&&\\ 
 Coviar mv& \textbf{69.84} & 70.27 & 70.91 & 70.34\\ 
 
 Refined mv& 69.57 & \textbf{70.49} & \textbf{71.51} & \textbf{70.52}\\
 \hline
\end{tabular}
\end{center}
\end{table}
However, we argue that it is more important to measure the performance when mv models work together with other models, such as I-frame subnet. Since the representation power of a shallow Resnet-18 is quite limited, it is more important to test how much other models can benefit from our proposed method. We measure the accuracy gain with the original I-frame subnet of Coviar, which is a Resnet-152 with average accuracy of 53.3\% and 87.7\% on Hmdb51 and UCF101. Results are shown in Table~\ref{tab:tab2}. On the Hmdb dataset, original Coviar's mv subnet provides 5\% accuracy gain, while our proposed mv refinement further improve the gain by 1\%. 
\begin{table}[h]
\begin{center}
 \caption{Accuracy Comparison MV + I-frame}
 \label{tab:tab2}
 \begin{tabular}{c c c c c} 
 \hline
& Split1 & Split2 & Split3 & Average \\ 
 \hline
 \textbf{Hmdb51} & & & & \\ 
 Coviar mv + I & 59.09 & 57.58 & \textbf{59.28} & 58.65\\ 
 
 Refined mv + I & \textbf{62.09} & \textbf{57.97} & 59.02 & \textbf{59.69}\\
 
  \textbf{UCF101} &&&&\\ 
 Coviar mv + I & 90.11 & \textbf{89.56} & 89.91 & 89.86 \\ 
 
 Refined mv + I & \textbf{90.43} & 89.29 & \textbf{90.12} & \textbf{89.94}\\
 \hline
\end{tabular}
\end{center}
\end{table}
Though 1\% might not seem to be a large increment, we further compare the refined mv subnet + I-frame subnet with Coviar's three subnet work fusions (mv + I-frame + residual(residual subnet is also a Resnet-18). Results are in Table~\ref{tab:tab3}.   On Hmdb51 dataset, the proposed method using only two subnets achieves better accuracy than original fusion of three subnets. As a result, we argue that the refined mv subnet provides more accuracy gains. And it is more efficient in that the computation of residual subnet could be saved. Detailed efficiency comparisons are discussed in the next section.
\begin{table}[h]
\begin{center}
 \caption{Comparison with Coviar's three-network fusion}
 \label{tab:tab3}
 \begin{tabular}{c c c c c} 
 \hline
& Split1 & Split2 & Split3 & Average \\ 
 \hline
 \textbf{Hmdb51} & & & & \\ 
 Refined mv + I & \textbf{62.09} & 57.97 & \textbf{59.02} & \textbf{59.69}\\ 
 
 Coviar mv+I+R & 60.4 & \textbf{58.2} & 58.7 & 59.1\\

 \hline
\end{tabular}
\end{center}
\end{table}
\subsection{Speed and Efficiency}

We evaluated the proposed method's efficiency on all test splits of Hmdb51 and UCF101, and results comparison with original Coviar are in Table~\ref{tab:tab4}. The units are sec/video. Hmdb51 has 6766 video clips with in total 632655 frames, while UCF101 has 13320 video clips with in total 2486290 frames. Hence it could be interpreted that the proposed method averagely process 254 frames/sec on hmdb51 and 486 frames/sec on UCF101 during test time, including video overheads.
\begin{table}[h]
\begin{center}
 \caption{Speed Comparison}
 \label{tab:tab4}
 \begin{tabular}{c c c c c} 
 \hline
sec/video & Split1 & Split2 & Split3 & Average \\
 \hline
 \textbf{Hmdb51} & & & & \\ 
 Coviar mv & 0.2959	& 0.2978 &	0.2995 & 0.2977\\ 
 
 Refined mv& 0.3580 & 0.3660 &	0.3823 & 0.3688\\
 
  \textbf{UCF101} &&&&\\ 
 Coviar mv& 0.3122	&0.3084	&0.3014	&0.3073 \\ 
 
 Refined mv& 0.3810	&0.3894	&0.3810	&0.3838\\
 \hline
\end{tabular}
\end{center}
\end{table}
\begin{figure*}
\centering
\includegraphics[height=8.5cm, width=14cm]{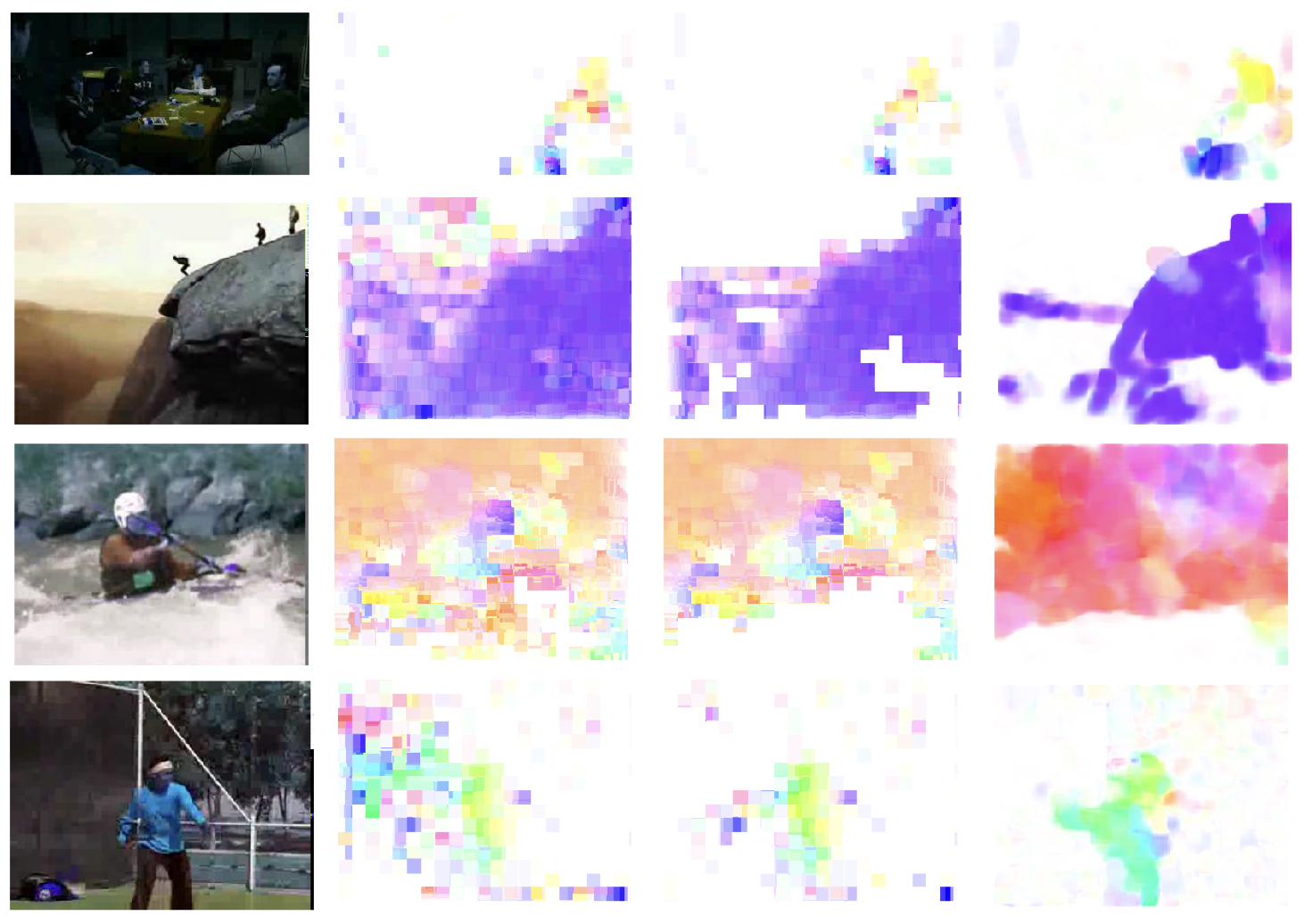}
\caption{More Results of the Proposed Method. Left to Right: Raw I-frame; Raw Motion Vector; The refined Motion Vector; Dense Optical Flow (Ground Truth). Videos from Hmdb51 dataset.}
\label{fig:final_res}
\vspace{-.8em}
\end{figure*}
Based on Table~\ref{tab:tab4}, we argue that the proposed method sees negligible efficiency degrades. One might argue that the drop from 0.3073 sec/video to 0.3838 sec/video is a large efficiency drop. However, as discussed in Section~\ref{sec:4.3}, the main purpose of the proposed method is to increase accuracy gain when working with other models. And take Coviar's I-frame subnet for example, the Resnet-152 takes 0.9984 sec/video on Hmdb51 and 0.9652 sec/video on UCF101. Comparing with that, the 0.08 sec/video efficiency drop is indeed negligible.

Besides, as compared in Section~\ref{sec:4.3}, the refined mv subnet improves I-frame subnets more than original Coviar's mv plus residual subnet does. Considering that, we could even remove the necessity for the residual subnet (which is as fast as original mv subnet, around 0.3 sec/video) and hence our refined method could even be faster than original Coviar. 

\subsection{Improvements with Network Depth}
One of the key requirements for CNN to learn complex visual features is sufficient network depths. The low-level convolutional layers learn fundamental features while deep-level convolutional layers learn complicated features. It is believed that the network capacity grows with network depth. Based on this intuition, we tested how much our refined motion vector could probably gained from deeper network architectures.

Instead of the original Resnet-18, we tried the refined mv subnet with Resnet-34. All pre-processing such as threshold the confidence map is the same as Section 3, and we trained 100 more epochs for the network to converge. The result is that on Hmdb51 Split1, the test accuracy further improves 1.5\% compared with using Resnet-18. And the speed decreases around 0.01 sec/video

By further increasing the network depths to, let say, Resnet-152, we could expect the mv subnet to perform as good as optical-flow stream in two-stream architectures. And there are many works attempting to approximate optical flows using very deep networks. However, the purpose of this paper is to develop a small subnet that could improve other network’s accuracy with negligible efficiency degrades.  And we keep Resnet-18 as our architecture, to allow machines with limited to no GPU memory to compute within a reasonable time.

\section{Conclusion}
Through this paper, we proposed a process of refining motion vectors to approximate optical flows and a model that leverages the refined motion vectors to perform action recognition. The proposed method achieved better results compared with original compressed video action recognition on both standard action recognition datasets, Hmdb51 and UCF101. More importantly, the proposed model could provide more accuracy gain when combined with other action recognition networks. The small model size generates few to no efficiency degrades, and makes it possible for machines with few GPU resources to process at a fast rate. 

\section*{Acknowledgement} 
The code implementations of this paper are modified based on the Github project pytorch-coviar. 

{\small
\bibliographystyle{ieee}
\bibliography{main}
}

\end{document}